\documentclass[11pt]{article}
\usepackage{epsfig}
\usepackage{cite}

\newcommand{\la}{$\Lambda$}
\newcommand{\al}{$\bar{\Lambda}$}
\newcommand{\lll}{$\Lambda(\bar{\Lambda})$}

\begin{document}

\begin{center}
\begin{Large}
{Measurements of $\Lambda$ and $\bar \Lambda$ polarization in longitudinally polarized proton-proton collisions at $\sqrt s$ =200 GeV at STAR}
\end{Large}

\vspace*{0.8cm}

Qinghua Xu for the STAR collaboration\\
\vspace*{0.3cm}
\footnotesize{{\it Nuclear Science Division, MS 70R0319, Lawrence Berkeley National Laboratory, Berkeley, \\CA 94720}  \&
\\{\it Department of Physics, Shandong University, Ji'nan, Shandong 250100, China}}
\end{center}

\vspace*{0.3cm}

\begin{abstract}
Preliminary results for the longitudinal polarization of $\Lambda$ and ${\bar \Lambda}$ hyperons
in longitudinally polarized proton-proton collisions at $\sqrt s$ = 200 GeV are presented. 
The $\Lambda$($\bar \Lambda$) candidates are reconstructed at mid-rapidity ($|\eta|<1$) 
with the time projection chamber of the STAR experiment at RHIC, using
0.5\,pb$^{-1}$ collected in 2003 and 2004 with beam polarizations of up to 45\%.
Their mean longitudinal momentum fraction $x_F$ is about 8 $\times$$10^{-3}$ and 
their mean transverse momentum $p_T$ is about 1.5 GeV.
The analysis uses asymmetries of counts for different spin states of the colliding proton beams and does not require detailed knowledge of the detector acceptance.
The preliminary $\Lambda$(${\bar \Lambda}$) polarization values are consistent with zero within their statistical uncertainties of 0.05.
\end{abstract}

\vspace*{0.6cm}

Hyperon polarization measurements in high energy polarized $pp$ collisions can give insight into the  spin transfer from a polarized quark to a polarized baryon~\cite{Florian98}.
Hyperon polarization can also shed light on the the polarized parton distribution functions~\cite{XL04}.
Anti-Lambda polarization provides sensitivity to $\Delta \bar s$~\cite{XLS05}.  

The $\Lambda$ and $\bar \Lambda$ have been studied widely because of their sizable production cross sections and self-analyzing decay properties in the decay channel $\Lambda \to p \pi^-$ ($\bar \Lambda \to \bar p \pi^+$) with a branching ratio of 64\%.
The polarization is usually extracted from the angular distribution of the decay (anti-)proton in the \lll\ rest frame:
\begin{equation}
\frac{dN}{d \cos{\theta}}=\frac{N_{tot}}{2}A(\cos \theta)(1+\alpha P
\cos{\theta}), 
\label{ideal}
\end{equation}
where $N_{tot}$ is the total number of \lll s,
$\alpha=+(-)0.642\pm0.013$ is \la(\al)~decay parameter~\cite{PDG}, 
$P$ is the \lll\ polarization, $\theta$ is the angle
between the (anti-)proton momentum in the \lll\ rest frame and the polarization direction, and $A(\cos\theta)$ is the detector acceptance.
In the longitudinal case the polarization direction is the moving direction of \lll.
In the notation above, we have omitted the dependence of the acceptance function on variables other than $\cos\theta$ for notational clarity.

\begin{figure}
\begin{center}
  \includegraphics[height=.25\textheight]{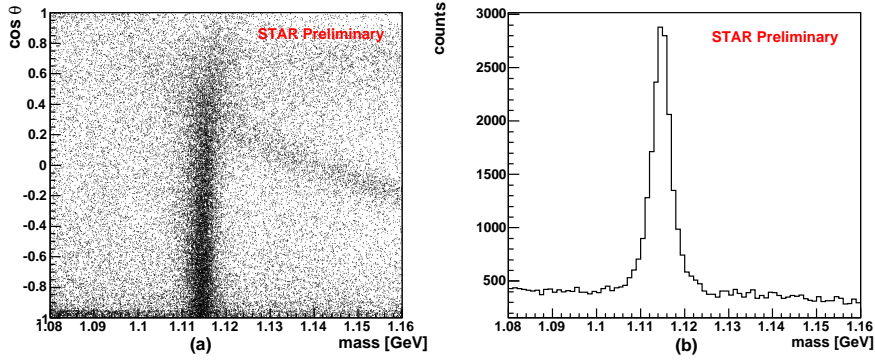}
  \caption{(a) The invariant mass versus $\cos \theta$ for the $\Lambda$ candidates after selections and (b) the projection on the mass axis. The invariant mass range of the \lll\ candidates is 1115$\pm$10 MeV.}
\label{mass}
\end{center}
\end{figure}

The data sample considered here consists of about 8M events, or about 0.5\,pb$^{-1}$, recorded in 2003 and 2004 by the Solenoid Tracker at RHIC (STAR) experiment~\cite{nim} in proton-proton collisions at $\sqrt s$ = 200 GeV with longitudinal beam polarizations up to 45\%.
The \lll\ reconstruction is made via decay $\Lambda \to p \pi^-$ ($\bar \Lambda \to \bar p \pi^+$).
The \lll\ candidates are reconstructed from two tracks with opposite curvature and a topology that is consistent 
with hyperon decay.
Selections on specific energy loss $dE/dx$ of protons and pions are made to reduce background.
Fig.\ref{mass}(a) shows the reconstructed invariant mass after selections versus $\cos \theta$ for a representative sample.
The signal is seen together with combinatoric background and a band originating from $K_s^0 \to \pi^+\pi^-$ decays.
The slight tendency of the reconstructed invariant mass to increase with $\cos \theta$ is caused by detector resolution.
Fig.\ref{mass}(b) shows the projection onto the mass axis.
To avoid $K_s^0$ being misidentified as \lll, events with $\cos(\theta) > -0.2$ are excluded from the analysis.

During longitudinally polarized running at RHIC proton beam bunches with positive and negative helicities are collided in four different combinations, $++$, $+-$, $-+$, and $--$.
Events recorded at STAR are sorted accordingly. 
Parity conservation in the QCD production of \la\ leads to the following relations for the \la\ polarization:
\begin{equation}
 P^{++}_\Lambda(\eta)=-P^{--}_\Lambda(\eta),~P^{+-}_\Lambda(\eta)=-P^{-+}_\Lambda(\eta), 
\label{symmetry}
\end{equation}
where $\eta$ denotes the pseudo-rapidity of the \la.
Similarly, $P^{+0}_\Lambda(\eta)=-P^{-0}_\Lambda(\eta)$ if the events samples can be combined to 
get a sample as if one beam were unpolarized (denoted by ``0'').
The \la\ polarization thus changes sign if the beam polarization(s) are reversed.
This is the same for \al.

The four beam spin patterns at RHIC and the above symmetry relations make it possible to extract the longitudinal \la\ polarization without detailed knowledge of the detector acceptance $A(\cos \theta)$ in Eq.~(\ref{ideal}).
Suppose one identifies two \la\ samples of equal size but with opposite (unknown) polarizations, 
$P_\Lambda$ and $-P_\Lambda$, and considers a small interval in the decay angle $\theta$, [$\theta_1,\theta_2$].
The asymmetry
\begin{equation}
 A_s=\frac{N(P_\Lambda)-N(-P_\Lambda)} {N(P_\Lambda)+N(-P_\Lambda)} 
\simeq \alpha P_\Lambda  \frac {\cos \theta_1 +\cos \theta_2}{2},
\label{pol}
\end{equation}
is proportional to $P_\Lambda$.
Here $N(P_\Lambda)=\int \frac{dN}{d\cos \theta} d\cos\theta$ denote counts with polarization $P_\Lambda$ 
and $N(-P_\Lambda)$ the counts with the opposite polarization.
The interval [$\theta_1,\theta_2$] is taken small enough so that the acceptance $A(\cos\theta)$ is a constant 
and cancels in the ratio.
Therefore, the \la~ polarization value can be extracted from the asymmetry $A_s$.
At RHIC the degree of beam polarization is measured with polarimeters \cite{polmeter} for 
each of the proton beams individually.
The relative luminosities for the four helicity combinations are measured with Beam Beam Counters\cite{BBC}.
These factors are taken into account in the extraction of the \la\ polarization through modifications to Eq.~
(\ref{pol}).
We have verified that the vertex distribution for the four helicity combinations, as well as several other distributions, have the same shape for the analyzed data samples so as to ensure that the detector acceptance $A(\cos\theta)$ may be taken equal for the spin-sorted samples.

\begin{figure}
\begin{center}
  \includegraphics[height=.20\textheight]{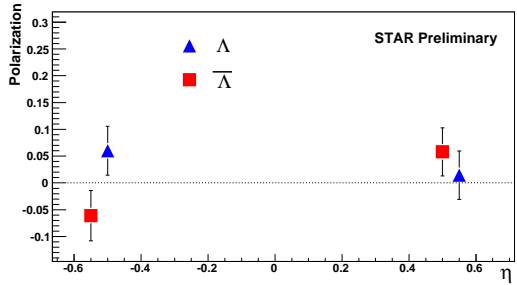}
\caption{Preliminary STAR results for $\Lambda$ and $\bar \Lambda$ longitudinal polarization in $pp$ collisions at $\sqrt s=200$ GeV for positive and negative pseudo-rapidities $\eta$.
The indicated uncertainties are statistical only.}
\label{dll}
\end{center}
\end{figure}

After data selections approximately 30K \la's and 27K \al's remain.
Their mean $x_F$ is about 0.008 and the mean $p_T$ is 1.5 GeV.  
Fig.\ref{dll} shows preliminary results for the \lll~polarization, 
with only one beam polarized, which is also referred to in the literature as the spin transfer $D_{LL}$.
Positive $\eta$ is taken along the moving direction of the polarized beam.
The data is split into positive $\eta$ and negative $\eta$, since the polarization may vary with $\eta$.
The indicated uncertainties are statistical only.
The systematic uncertainty from the relative luminosity ratio $R$ is estimated to be about 0.01.
The uncertainty from the beam polarization measurement is about 20\%.
Studies of possible bias caused by triggering are in progress.

The observed $\Lambda$ and $\bar \Lambda$ polarizations are consistent with zero 
within statistical uncertainties of 0.05, 
as qualitatively expected from the small average $x_F$ and $p_T$. 
Quantitative theoretical predictions exist for transverse momenta $p_T$ larger than 8\,GeV,
where perturbative calculations may be expected to hold.
It should be noted that at lower pT, for which preliminary data exist\cite{mheinz}, 
the cross section is not well described by current  
pQCD calculations.
The data sample recorded in 2005 is about 2 times larger than that collected in 2003 and 2004.

 This work was supported in part by the U.S. Department of Energy.

\end{document}